\documentstyle[12pt,graphicx,color,
epsfig,cite,amsmath]{article}
\def\baselinestretch{1.3}
\newcommand{\ba}{\begin{array}}
\newcommand{\ea}{\end{array}}
\newcommand{\bd}{\begin{displaymath}}
\newcommand{\ed}{\end{displaymath}}
\newcommand{\be}{\begin{equation}}
\newcommand{\ee}{\end{equation}}
\newcommand{\bea}{\begin{eqnarray}}
\newcommand{\eea}{\end{eqnarray}}
\newcommand{\bei}{\begin{itemize}}
\newcommand{\eei}{\end{itemize}}

\def\a{\alpha}

\def\b{\beta}

\def\q2 {q^2}

\def\bt{\begin{table}}
\def\et{\end{table}}

\def\a{\alpha}

\def\b{\beta}

\def\q2 {q^2}

\def\bt{\begin{table}}
\def\et{\end{table}}

\catcode`@=11 
\def \gsim{\mathrel{\mathpalette\@versim>}}
\def \lsim{\mathrel{\mathpalette\@versim<}}
\def \@versim#1#2{\lower0.4ex\vbox{\baselineskip\z@skip\lineskip\z@skip
     \lineskiplimit\z@\ialign{$\m@th#1\hfil##\hfil$%
     \crcr#2\crcr\sim\crcr}}}
\catcode`@=12 

\def\wt{\widetilde}
\def\te{\tilde e}

\def\tu{\tilde u}

\def\tb{\tilde b}

\def\tst{\tilde t}

\def\tg{\tilde g}

\def \lspone{\wt\chi_1^0}
\def \mlspone{m_{\lspone}}
\def \lsptwo{\wt\chi_2^0}
\def \mlsptwo{m_{\lsptwo}}
\def \lspthree{\wt\chi_3^0}
\def \mlspthree{m_{\lspthree}}
\def \lspfour{\wt\chi_4^0}
\def \mlspfour{m_{\lspfour}}

\def\issue(#1,#2,#3){{\bf #1}, #2 (#3)}

\def\PREP(#1,#2,#3){Phys.\ Rep. \issue(#1,#2,#3)}

\begin{document}

\begin{flushright}
OSU-HEP-11-1\\\
 UCRHEP-T502
\end{flushright}

\begin{center}
{\large \bf Supersymmetry Signals at the LHC under the most favorable SUGRA
scenario}\\[15mm]
Subhaditya Bhattacharya\footnote{E-mail: subhab@ucr.edu} and
S. Nandi\footnote{E-mail: s.nandi@okstate.edu}\\[3 mm]
 1.{\em Department of Physics and Astronomy, \\
     University Of California, Riverside, CA 92501, USA}\\[2 mm]
2.{\em Department of Physics, Oklahoma State University, and
Oklahoma Center for High Energy Physics, Stillwater, OK 74078, USA}
\\[20mm]
\end{center}

\begin{abstract}
\noindent

Given that it will take quite some time for the Large Hadron Collider (LHC)
to reach its desired luminosity, it is important to investigate the most
favorable scenario in which supersymmetry (SUSY) may be discovered
at the early runs at the LHC. Our aim in this work is to find such a
scenario within the gravity mediated SUSY breaking (SUGRA) framework
and select a class of final states that warrant a discovery at the
very early runs of the LHC. It turns out that such a situation can be associated
with a scenario where gluinos are sufficiently light and  so are the
third generation scalars while the first two family scalars are heavy.
We find that this can be achieved from a
high-scale set-up with scalar mass non-universality in the third family and
gaugino mass non-universality with $M_3 < M_1, M_2$.
We show that the final state channels which
are most favorable in such a region of parameter space are
 $4b+{E_{T}}\!\!\!\!/$, $4b+\ell+{E_{T}}\!\!\!\!/$ and
$2b+2\ell+{E_{T}}\!\!\!\!/$. We also justify our claim by comparing
the results with a minimal supergravity (mSUGRA) scenario with
similar gluino mass.

\vskip 15pt
\noindent
\end{abstract}



\newpage

\setcounter{footnote}{0}

\def\baselinestretch{1.5}

\section{Introduction}

With the Large Hadron Collider (LHC) already collected data for
the run with a center-of-mass energy $E_{CM}$=7 TeV, the
search for physics beyond the Standard Model has reached a new
height of excitement. Supersymmetry (SUSY) has been one of the most
popular scenarios in this category, due to its attractive
theoretical framework and variety of phenomenological features it
incorporates. Apart from stabilizing the Higgs sector, it also
provides with a cold dark matter candidate in form of the lightest
SUSY particle (LSP) with $R$-parity (${(-1)}^{3B+L+2S}$)
conservation \cite{SUSY,KaneKingRev} .

Out of different SUSY-breaking models, the most popular one is the minimal supergravity or
mSUGRA \cite{msugra}. Here all the SUSY-breaking parameters are derived from
four and half parameters, namely, the universal gaugino mass
($M_{1/2}$), the universal scalar mass ($m_0$), the universal
trilinear coupling ($A_0$) all at the GUT scale, $\tan\beta$, the
ratio of the vacuam expectation values (vev) of the two Higgses and
the sign of SUSY-conserving Higgsino mass parameter $\mu$. However,
proposals have been made to go beyond the universality of scalar
\cite{ucddOct2008,berez,Nath:1997qm,Cerdeno:2004zj,ellis-all,baer-all-non,so102,Datta:1999uh,BM-SB-AD2}
 and gaugino masses \cite{nonunigaugino,BM-AD-SB} within the SUGRA framework
itself, which are strongly motivated from supersymmetric Grand Unified Theory
(SUSY-GUT) and low energy phenomenological constraints, for example, suppression of the
flavor changing neutral current (FCNC) or CP-violation.

 With the prospect of LHC running at $E_{CM}$= 14 TeV,
 it is quite likely to shed light upon SUSY if it exists at
the TeV scale. Whether this is possible, and if at all, then when
and how, depend strictly on the SUSY spectra that nature has chosen
for us. The most likely manifestation of SUSY will be the excess in
certain final state channels over the standard model (SM)
background fluctuations.

Following this, the main trends of the phenomenological studies in
search of SUSY in context of the LHC have been directed either to a
model-based approach where one has a specific SUSY model under consideration
and then to study the consequences in a collider, or to interpret
the measured excess in signals to extract the SUSY- parameters such as mass,
spin etc. using kinematic variables and/or distributions \cite{mass}. There have been 
efforts to study the so-called `LHC inverse problem'  \cite{ArkaniHamed:2005px} as well. 
While all these are very important and of absolute necessity at the
same time, perhaps a combined knowledge of all such studies may provide us
with the most useful hint to unravel SUSY at the upcoming collider
experiment.

In this work, we try to address a relatively simple question - what
is the most favorable supersymmetric spectra allowed by all the
search limits and low energy constraints that can leave its imprint
during the early run of the LHC and correspondingly what are the
final states that are favorable for such a spectra. Our approach
here is pragmatic, particularly from the experimental point of view,
as well as, somewhat model independent to start with. We of course,
associate our proposed spectra to a SUGRA pattern to justify our claim.

The answer to the question raised here, may look very simple at the
first sight: have the minimum possible values of all SUSY particle
masses allowed by current bounds.
However, such a choice may not be consistent with a SUSY breaking
scenario. We argue that a spectra with light gluinos, light third family scalars,
but relatively heavy first two families (that can be derived from a
high-scale non-universal SUGRA pattern), is one of the most favorable cases
for the SUSY to be discovered at the early LHC runs. It is
even more true when we look for the final states
in form of $4b+{E_{T}}\!\!\!\!/$, $4b+\ell+{E_{T}}\!\!\!\!/$ and
$2b+2\ell+{E_{T}}\!\!\!\!/$ as these arises from the decays of the dominant
SUSY production processes in such a region of parameter space
and  have little contribution from the SM background.
To prescribe these as the golden modes for early SUSY discovery,
we of course, assume that the machine is tuned properly to measure missing
energy and tag $b$-jets. This is also justified because the main
characteristic of SUSY-signature is high missing energy which is
carried away by the LSP in the $R$-parity conserving framework.

We see that, in order to keep the third family
squarks much lighter than the first two, in our proposed benchmark scenario
and to associate them with a SUGRA pattern, we require a scalar
non-universality at high scale. On the other hand, to keep gluinos light and low-lying
electroweak gauginos above the LEP limit at the same time, we
require a gaugino mass non-universality with a heirarchy of
$M_3 < M_1, M_2$ at the high scale. The gaugino mass non-universality
of the form $M_3 < M_1, M_2$ is
achievable within the framework of the SUSY-GUT \cite{nonunigaugino,BM-AD-SB}, while a scalar
non-universality of such kind can be motivated from string-inspired
models with flavor dependent couplings to the modular fields \cite{3rd1,3rd2}.
To justify our claim, we also compare our
results with a mSUGRA parameter point that gives similar gluino
mass.

Our paper is organized as follows. In Section 2, we discuss the model
under consideration and the selected benchmark points chosen for
further studies concerning collider signatures. In Section 3, we
discuss the final states that we look for, the details of the
collider simulation strategy adopted here and the numerical results
obtained from this analysis. We conclude in Section 4.

\section {Model, Formalism and Benchmark Points}

In this section we first advocate the favorable SUSY spectra under
consideration. We also motivate the final state signals in which
such a parameter space can be better observed. Then we show that
such a spectra can be associated to a high-scale non-universal
gaugino and scalar mass set-up in a GUT- based SUGRA framework. At
the end of this section, we discuss the benchmark points chosen for
studying the collider signals.

It is easy to appreciate that a
scenario which claims to be favorable for discovery should have
very light gluinos simply because it
has a very high production rate at the LHC (gluinos are colour octets and
the main production through gluons gets enhanced because of high
gluon flux at this energy regime). What about the
squarks ? If these are light, they also compete in production rate
with the gluinos, given the fact that they also have strong interaction and comes with
 three families. In a generic sense, then light squarks are also favorable
to add to the SUSY-final states. We need to remember also that
the gluino dominantly decays through
the on-shell squark-quark giving rise to more likely a jet-rich
final state if first two generation squarks participate in the decay
chain. The jetty final states have a large SM backgrounds coming
from QCD processes. Associated leptons may arise in such cases, but,
mostly from the decays of the electroweak gauginos that appear in
the cascades in such situations. Increasing the possibility of final
state leptons is better to see SUSY signals over
the background fluctuations in a hadron collider machine.
Although the situation with gluinos and squarks all being light,
can in principle, warrant a discovery at the early run of
LHC by suitable choice of event selection criteria, the question is can we think
of something more distinctive? Actually we
can exploit the fact that heavy quarks, namely the top and the
bottom have the property of decaying to leptons or get identified by
b-tagging itself. So, if the squarks belonging to the the first two
generations be very heavy, then the gluinos will decay dominantly through
sbottom-bottom or stop-top. While, with a similar sbottom and stop mass, the
gluino is more likely to decay through sbottom-bottom simply because of the phase space.
Both cases give rise to bottom quark
rich final states. Then, if we look for the final states with
multiple $b$s, namely, $4b+{E_{T}}\!\!\!\!/$,
$4b+\ell+{E_{T}}\!\!\!\!/$ or $2b+2\ell+{E_{T}}\!\!\!\!/$, they
capture the gluinos decaying through sbottom-bottom or stop-top,
with the  sbottom decaying through bottom-neutralino or a
top-chargino and the stop decaying to top-neutralino or bottom-chargino.
In each case, we expect bottom reach final state. Also contribution from stop and sbottom pair productions contribute to such kind of final
states.

 We discuss the final states with the cuts in details in
Section 3.

From the model point of view, it also turns out
that both FCNC and CP-violation constraints may be best tackled if
one assumes the first two generations of scalars to be multi-TeV and
(quasi-)degenerate in masses.~\footnote{We remind the reader that
satisfying constraints imposed by electric dipole moments of
electron and neutron would require very large scalar masses if we
like to have finite values for the CP-violating SUSY phases.} This
so-called `inverted hierarchy' is favored from low-energy
constraints as well.

A scalar mass hierarchy of this sort can be achieved from a high-scale
non-universality in first two family scalars with the third one. More
specifically, if the squark masses get generated from two different
uncorrelated mass parameters namely, ${m_0}^3$ and ${m_0}^{1,2}$
respectively for the third generation and first two generations at
the high scale, then choosing a high value of ${m_0}^{1,2}$ and a
small value of ${m_0}^3$, given a particular value of high scale
gaugino mass parameter can yield a spectra of the pattern discussed
above. Although, it is a phenomenological framework that we discuss
here, it can be motivated from string inspired models with flavor
dependent couplings to the modulaii fields \cite{3rd1,3rd2}. We would like to note
that such a set up has been discussed early in some articles for
studying the collider signature in context of the LHC \cite{ucddOct2008, BM-SB-AD2}, but as
mentioned early, not specifically in this context.

In order to have the gluinos as light as possible
and at the same time the low-lying charginos
and neutralinos above the threshold, we require a non-universality in
gaugino masses to accommodate it in a SUGRA
pattern. The hierarchy required at the high-scale is $M_3 < M_1,
M_2$. This can be generated within the framework of
SUSY-GUT with an underlying $SU(5)$ or $SO(10)$ gauge symmetry 
\cite{nonunigaugino, BM-AD-SB}.
Often one can incorporate a dimension five operator in the
non-trivial extension of the gauge kinetic function $f_{\a
\b}(\Phi^{j})$ in terms of the non-singlet chiral superfields
$\Phi^N$ of the form
\bea
 Re f_{\a \b}(\phi)F_{\mu \nu}^{\a}F^{\beta \mu \nu}= \frac {\eta (\Phi^s)}
{M}Tr(F_{\mu \nu}\Phi^N F^{\mu \nu})
\eea
\noindent
where $\Phi^N$ belongs to the symmetric product of the adjoint representation of
the underlying gauge group as\\
\begin{eqnarray}
SU(5):    &  (24\times 24)_{symm} = 1+24+75+200 \\ \nonumber
SO(10):   &   (45\times 45)_{symm}=1+54+210+770
\end{eqnarray}
Gaugino masses become non-universal if the Higgses responsible for
the GUT-breaking, belong to the possible non-singlet
representations, unlike the minimal supergravity (mSUGRA) framework
\cite{nonunigaugino,BM-AD-SB}. Interestingly, the representations $75$ and $200$ belonging to
$SU(5)$ or $770$ \footnote{For breaking through $770$, we quote
the result, when it breaks through the Pati-Salam gauge group
$G_{422D}$ ($SU(4)_C \times SU(2)_L\times SU(2)_R$ with even
D-parity and assumed to break at the GUT scale itself.} of $SO(10)$
break the GUT group to the SM, we obtain the required hierarchy of
 $M_3 < M_1, M_2$ at the GUT scale.
We tabulate the non-universal gaugino mass ratios for these
representations in Table 1. A linear combination of these
non-singlet representations with the singlet one can yield the exact
hierarchy that we use in the benchmark points.
We would also like to mention that such a gaugino mass hierarchy is
also supported from the dark- matter consideration, as the immediate
effect of having a smaller $M_3$ yields a smaller $\mu$ after the
RGE running, yielding a more Higgsino like lightest neutralino,
which has much efficient annihilation rate to yield a consistent cold dark matter relic density.

\begin{table}
\begin{center}
\begin{tabular}{|c|c|}
\hline
\hline
 Representation & $M_{3}:M_{2}:M_{1}$ at $M_{GUT}$ \\
\hline
{\bf 75} of $SU(5)$ & 1:3:(-5) \\
\hline
{\bf 200} of $SU(5)$ & 1:2:10 \\
\hline
{\bf 770} of $SO(10)$: {$H \rightarrow SU(4) \times SU(2) \times SU(2)$} &
1:(2.5):(1.9) \\
\hline
\hline
\end {tabular}
\end{center}
{\caption {\em Non-universal gaugino mass ratios for different non-singlet
representations belonging to $SU(5)$ or $SO(10)$ GUT-group that gives rise
to the hierarchy of $M_3 < M_1, M_2$ at the GUT scale.}}
\end{table}

Now we discuss the benchmark points chosen for
the study of collider signature from this framework. We stick to
have a high-scale universal $A_0$ set to zero. For all the points
$sgn(\mu)$ has been taken to be positive and the Higgs mass
parameters has been set equal to the third generation scalar masses
$m_{H_{u}}^2=m_{H_{d}}^2={{m_0}^3}^2$. The values of  $\tan \beta$ has been chosen
such that it satisfies the experimental constraint branching ratio for the $b \longrightarrow s \gamma$ \cite{bsg-recent} which
at the $3 \sigma$ level is
\begin{equation}
2.77 \times 10^{-4} < Br (b \rightarrow s \gamma) < 4.33 \times 10^{-4}.
\label{bsgammalimits}
\end{equation}

Parameters are fine-tuned in a way that it gives a correct cold dark matter
relic abundance. In cases it is smaller than the WMAP data \cite{WMAPdata},
which at $3 \sigma$ is
\begin{equation}
0.091 < \Omega_{CDM}h^2 < 0.128 \ ,
\label{relicdensity}
\end{equation}
where $\Omega_{CDM}$ is the dark matter
relic density in units of the critical
density and $h=0.71\pm0.026$ is the reduced Hubble constant
(namely, in units of
$100 \ \rm km \ \rm s^{-1}\ \rm Mpc^{-1}$).

This leaves us with the opportunity for some other dark-matter component.
As we mentioned earlier we compare our result with a mSUGRA point,
which has a similar gluino mass. We denote the benchmark points as
BP1, BP2 and BP3, while the mSUGRA point is denoted as MSG1. We tabulate the
high-scale and low scale parameters in Table 2. BP2 has the lightest gluino among these and BP3 has the heaviest one.

Note that for BP3, we do not need a gaugino mass
non-universality to keep the lighter chargino, neutralinos to rise
above the experimental bound.

\begin{table}
\begin{center}
\begin{tabular}[ht]{|c|c|c|c|c|}
\hline
\hline
parameter & BP1 & BP2 & BP3 & MSG1\\
\hline
\hline
$\tan\beta$ &17  &12 &23& 5\\
$(M_3,M_2,M_1$) &(140,180,180)  &(125,250,250) & (153,153,153) & (200,200,200)\\
$({m_0}^3,{m_0}^{1,2})$ &(160,1000)  &(160,1000) & (160,1000)& (70,70)\\
$A_0$ & 0  & 0 & 0&0\\
$sgn(\mu)$ &+  &+ &+&+\\
\hline
\hline
$\mu$ & 166  & 123 & 190 & 285 \\
$m_{\tg}$ & 389  & 349 & 421 & 495 \\
$m_{\tu_L}$ & 1037  & 1034   & 1043 & 457\\
$m_{\tst_1}$ &  187 & 142 & 217 & 324\\
$m_{\tst_2}$ &  364 & 348 & 382 & 501 \\
$m_{\tb_1}$ & 268 & 253 & 285 & 427 \\
$m_{\tb_2}$ & 302 & 273 & 330 & 447 \\
$m_{\te_L}$ & 1005 & 1012  & 1003 & 159  \\
$m_{{\tilde \tau}_1}$ & 157  & 178 & 138 & 107 \\
$m_{{\tilde\chi_1}^{\pm}}$ & 104  & 98  & 98.5 & 132 \\
$m_{{\tilde\chi_2}^{\pm}}$ & 223 & 244 & 233 & 317\\
$\mlspfour$ & 225 & 245 & 232 & 319\\
$\mlspthree$ & 178 & 135 & 203 & 291\\
$\mlsptwo$ & 110 & 125 & 101 & 134\\
$\mlspone$ & 64 & 71 & 56 & 73\\
\hline
\hline
$\Omega_{\tilde\chi_1}h^2$& 0.08 & 0.035 & 0.128 & 0.128\\
$BF(b\to s\gamma)$ & $3.04\times 10^{-4}$ & $4.3\times 10^{-4}$
& $4.27\times 10^{-4}$ & $3.09\times 10^{-4}$\\
\hline
\hline
\end{tabular}
\end{center}
\caption {\em : Benchmark points BP1, BP2, BP3 and MSG1(masses are in GeV). The
first five parameters define the model, while the rest are low scale
prediction.}
\end{table}

For renormalization group equation RGE, we use the code {\tt SuSpect v2.3} \cite{suspect}
and stick to two-loop RGE with
radiative corrections to the gauginos and squarks. We use full one loop and
dominant two loop corrections for the Higgs mass. The low scale value of the strong
coupling constant has been chosen at
${\alpha_3 (M_{Z})}^{\overline{MS}}= 0.1172$. We ensure radiative electroweak
symmetry breaking and the electroweak symmetry breaking scale has been set
at $\sqrt{m_{\tilde{t_{L}}}m_{\tilde{t_{R}}}}$, the default value in the code
{\tt SuSpect}. We compute the cold dark matter relic density with the code {\tt microOmega v2.0}
\cite{micromegas}.


\section {Collider Simulation and Results}

We now discuss the collider signatures of the benchmark points
advocated in the preceding section.

We first discuss the strategy for the simulation which includes the
final state observables and the cuts employed therein. In the next subsection
we discuss the numerical results obtained from this analysis.
\subsection{Strategy for Simulation}
The spectrum generated by {\tt SuSpect} v2.3 as described in the earlier
section, at the benchmark points are fed into the event generator
{\tt Pythia} 6.4.16 \cite{Pythia} by {\tt SLHA} interface \cite{sLHA}
for the simulation of $pp$ collision with centre of mass energy 7 TeV and
14 TeV.

We have used {\tt CTEQ5L} \cite{CTEQ} parton distribution functions,
the QCD renormalization and factorization scales being
both set at the subprocess centre-of-mass energy $\sqrt{\hat{s}}$.
All possible SUSY processes and decay chains consistent
with conserved $R$-parity have been kept open. We have kept
initial and final state radiations on. The effect of multiple
interactions has been neglected. However, we take
hadronization into account using the fragmentation functions
inbuilt in {\tt Pythia}.

As we have mentioned earlier that the spectra we have chosen for study is
best discovered with a $b$-reach final state. The final states studied here are:

\begin{itemize}
\item $(4b)$ :
$4b~+ X~ + {E_{T}}\!\!\!\!/$

\item $(4b+l)$ :
$4b~+ \ell~ + {E_{T}}\!\!\!\!/$

\item $(2b~+2\ell)$:
$2b~ + 2\ell~ + {E_{T}}\!\!\!\!/$

\end{itemize}

\noindent
where $\ell$ stands for final state isolated electrons and or muons,
${E_{T}}\!\!\!\!/$ depicts the missing energy, $X$ indicates any associated
lepton or jet production.

We will discuss these objects in details, that constitute the final state
observables. The nomenclature assigned to the final state events in parentheses
will be referred in the following text.

All the charged particles with transverse momentum, $p_T~>$ 0.5 GeV\footnote
{This is specifically for ATLAS, while for CMS, $p_T ~>$ 1 GeV is used.}
that are produced in a collider, are detected due to strong B-field within a
pseudorapidity range $|\eta|<5$, excepting for the muons where the range is
$|\eta|<2.5$, due to the characteristics of the muon chamber.
Experimentally, the main 'physics objects' that are reconstructed in a
collider, are categorized as follows:
\begin{itemize}
\item Isolated leptons identified from electrons and muons
\item Hadronic Jets formed after identifying isolated leptons
\item Unclustered Energy made of calorimeter clusters with $p_T~>$ 0.5 GeV
(ATLAS) and $|\eta|<5$, not associated to any of the above types of
high-$E_T$ objects (jets or isolated leptons).
\end{itemize}
Below we discuss the 'physics objects' described above in details.
\bei
\item {\em Isolated leptons} ($iso~\ell$):
\eei
Isolated leptons are identified as electrons and muons with $p_T>$ 10 GeV
and $|\eta|<$2.5. An isolated lepton should have lepton-lepton separation
${\bigtriangleup R}_{\ell\ell}~ \geq $0.2, lepton-jet separation
(jets with $E_T >$ 20 GeV) ${\bigtriangleup R}_{{\ell}j}~ \geq 0.4$,
the energy deposit $\sum {E_{T}}$ due to low-$E_T$ hadron activity around a
lepton within $\bigtriangleup R~ \leq 0.2$ of the lepton axis should be
$\leq$ 10 GeV, where  $\bigtriangleup R = \sqrt {{\bigtriangleup \eta}^2
+ {\bigtriangleup \phi}^2}$ is the separation in pseudo rapidity and
azimuthal angle plane. The smearing functions of isolated electrons, photons
and muons are described below.

\bei
\item {\em Jets} ($jet$):
\eei Although we do not use any explicit jet veto in our specific
final states excepting for the b-tagging we would like to discuss
the jets as they are one of the important ingredients of the final
state observables in a collider. Jets are formed with all the final
state particles after removing the isolated leptons from the list
with {\tt PYCELL}, an inbuilt cluster routine in {\tt Pythia}. The
detector is assumed to stretch within the pseudorapidity range
$|\eta|$ from -5 to +5 and is segmented in 100 pseudorapidity
($\eta$) bins and 64 azimuthal ($\phi$) bins. The minimum $E_T$ of
each cell is considered as 0.5 GeV, while the minimum $E_T$ for a
cell to act as a jet initiator is taken as 2 GeV. All the partons
within $\bigtriangleup R$=0.4 from the jet initiator cell is
considered for the jet formation and the minimum $\sum_{parton}
{E_{T}}^{jet}$ for a collected cell to be considered as a jet is
taken to be 20 GeV. We have used the smearing function and
parameters for jets that are used in {\tt PYCELL} in {\tt Pythia}.

\bei
\item {\em b-jets}:
\eei
We identify partonic $b$ jets by simple $b$-tagging algorithm with
efficiency of $\epsilon_b = 0.5$ for $p_T >$ 40 GeV and $|\eta| <$ 2.5 \cite{b-tagging-ref}.

\bei
\item {\em Unclustered Objects} ($Unc.O$):
\eei
 As has been mentioned earlier, all the other final state
particles, which are not isolated leptons and separated from jets by
$\bigtriangleup R \ge$0.4 are considered as unclustered objects.
This clearly means all the particles (electron/photon/muon) with
$0.5< E_T< 10$GeV and $|\eta|< 5$ (for muon-like track $|\eta|<
2.5$) and jets with $0.5< E_T< 20$GeV and $|\eta|< 5$, which are
detected at the detector, are considered as unclustered objects.

Once we have identified the 'physics objects' as described above, we sum
vectorially the x and y components of the momenta separately for all visible
objects to form visible transverse momentum $(p_T)_{vis}$,
\bea
(p_T)_{vis}=\sqrt{(\sum p_x)^2+(\sum p_y)^2}
\eea
where, $\sum p_x =\sum (p_x)_{iso~\ell}+\sum (p_x)_{jet}+\sum (p_x)_{Unc.O}$
and similarly for $\sum p_y$.
We identify the negative of the $(p_T)_{vis}$ as missing energy
$E_{T}\!\!\!\!/$:
\bea
E_{T}\!\!\!\!/ = -(p_T)_{vis}
\eea

Finally the selection cuts that are used in our analysis are as follows:

\begin{itemize}
\item  Missing transverse energy $E_{T}\!\!\!\!/$ $\geq ~100$ GeV.

\item ${p_{T}}^\ell ~\ge ~20$ GeV for all isolated leptons.

\item ${E_{T}}^{b-jet} ~\geq ~40$ GeV and $|{\eta}_{jet}| ~\le ~2.5$

\end {itemize}

Interestingly the channels we study, specifically the ones associated with
$4b$-s, do not have any dominant SM background at all. Although, the $4b$
production at LHC is significantly high, it is expected to get down with
the high missing energy associated with the final states. For the case with
$2b$, however, $t\bar t$ production gives the most serious
background. We generate SM events in {\tt Pythia} for the same final
states with same cuts. Our cuts and event selections have been motivated by \cite{Sanjoy}.

Studies in similar direction have also been discussed in references \cite{Recent}.

\subsection{Numerical Results}

Here we tabulate the event rates obtained for different channels at
different benchmark points advocated above. 

\begin{table}[!ht]
\begin{center}
\begin{tabular}{|c|r|r|r|r|}
\hline
Model Points & \multicolumn{1}{c|}{Total} &
  \multicolumn{1}{c|}{$\tilde g \tilde g$} &
\multicolumn{1}{c|}{$\tilde t_1 {\tilde t_1}^*$} \\
\hline
\hline
 {\bf BP1} & 141.1 & 50.5 & 45.3 \\
\hline
 {\bf BP2} & 315.6 & 95 & 164.5 \\
\hline
 {\bf BP3} & 100.6 & 33 & 23.6 \\
\hline
 {\bf MSG1} & 93.7 & 12 & 2.5 \\
\hline
\hline
\end {tabular}
\end{center}
\caption {\em  Total supersymmetric particle production cross-sections
(in pb) as well as some leading contributions for each of the
benchmark points. $E_{CM}$= 14 TeV.}
\end{table}

\begin{table}[!ht]
\begin{center}
\begin{tabular}{|c|r|r|r|r|}
\hline
Model Points & \multicolumn{1}{c|}{Total} &
  \multicolumn{1}{c|}{$\tilde g \tilde g$} &
\multicolumn{1}{c|}{$\tilde t_1 {\tilde t_1}^*$} \\
\hline
\hline
 {\bf BP1} & 19.9 & 3.9 & 6.0 \\
\hline
 {\bf BP2} & 48.1 & 8.1 & 26.0 \\
\hline
 {\bf BP3} & 16.0  & 2.2  & 2.6 \\
\hline
 {\bf MSG1} & 12.4 & 0.6 & 0.2 \\
\hline
\hline
\end {tabular}
\end{center}
\caption {\em  Total supersymmetric particle production cross-sections
(in pb) as well as some leading contributions for each of the
benchmark points. $E_{CM}$= 7 TeV.}
\end{table}

In Table 3 and Table 4, respectively for $E_{CM}$=14 TeV and 7 TeV, we note the total $2\rightarrow 2$ supersymmetric
production cross-section, as well as the dominant processes, namely
gluino and $\tilde{t_1}$ pair production. We can see BP2 has the highest production cross-section and then comes BP1, BP3 and MSG1 respectively. This can be easily attributed to the mass hierarchy of the SUSY mass spectra chosen for these benchmark points. For BP2, the stop pair production cross-section is even more than the gluino production at 14 TeV. Cross-sections for all the benchmark points decrease significantly at 7 TeV. Most interestingly, gluino production cross-section decreases significantly at 7 TeV for all the benchmark points. This is understandable as the gluon flux decreases at 7 TeV to a large extent which reduces the gluino production. Hence, for BP1, BP2 and BP3 the stop production cross-section is more than the gluinos at 7 TeV.

We also point out that for BP1, gluino dominantly decays to sbottom-bottom, while
sbottom dominantly decays to bottom neutralino. The stop1 in this case, decays dominantly 
to bottom-chargino. This remains the case for BP2 and BP3 as well. However, for the mSUGRA point, gluino decay is equally divided among all the squarks. Hence, we can easily see that  MSG1 lags far behind in channels advocated here, except for the $2b~ + 2\ell~ + {E_{T}}\!\!\!\!/$ channel. This is clear from Table 5 and 6, where we note the cross-sections for individual channels at $E_{CM}$= 14 TeV and $E_{CM}$= 7 TeV respectively. This is also due to the production cross-sections as noted earlier.

\begin{table}
\begin{center}
\begin{tabular}{|c|c|c|c|c|c|}
\hline
Benchmark Points & $\sigma_{4b}$ & $\sigma_{4bl}$ &$\sigma_{2b2l}$\\
\hline
\hline
BP1 & 1.50 & 0.15 & 0.43 \\
\hline
BP2 & 1.23 & 0.16 & 0.46 \\
\hline
BP3 & 1.17 & 0.17 & 0.33 \\
\hline
MSG1 & 0.04 & 0.01 & 0.40 \\
\hline
ttbar & 0.00 & 0.00 & 0.26 \\
\hline
\hline
\end {tabular}
\end{center}
\vspace{0.2cm}
{\caption{\em Event-rates (pb) in multichannels at the chosen
benchmark points for $E_{CM}$= 14 TeV. {\tt CTEQ5L} pdfset was
used. Factorization and Renormalization scale has been set to
$\mu_F=\mu_R=\sqrt{\hat s}$, subprocess centre of mass energy.}}
\end{table}

From Table 5, we see that the number of events in channel $4b~+ X~ + {E_{T}}\!\!\!\!/$ is as large as 1500 even for luminosity 1$fb^{-1}$ for 14 TeV for BP1. The number of $4b~+ \ell~ + {E_{T}}\!\!\!\!/$ events is also as large as 150 for the same luminosity at this point. For BP2 and BP3, these decrease, but still they appear with events of same order of magnitude. Contribution from $t\bar{t}$ background in these channels are almost zero. Hence, these are really promising channels to discover SUSY at the LHC. The channel $2b~ + 2\ell~ + {E_{T}}\!\!\!\!/$ although has a contribution from $t\bar{t}$ background, the significance ($S/\sqrt{B}$, where $S$ and $B$ signify signal and background events) for BP1 is as large as 26 at 1$fb^{-1}$. For BP2 and BP3 the significance is almost of the same order, which is sufficiently over the discovery limit. We also note that, as expected, MSG1 although has large number of events to be seen at this energy, but significantly smaller than the non-universal benchmark points particularly in channels  $4b~+ X~ + {E_{T}}\!\!\!\!/$ and  $4b~+ \ell~ + {E_{T}}\!\!\!\!/$.

At 7 TeV, although the number of events decrease significantly, still number of $4b~+ X~ + {E_{T}}\!\!\!\!/$ events for BP1 is as large as 115 at 1$fb^{-1}$. All the features discussed for 14 TeV, remains the same for 7 TeV as well. This indeed warrants our claim for an early SUSY discovery for the model and benchmark points advocated in our analysis.

\begin{table}
\begin{center}
\begin{tabular}{|c|c|c|c|c|c|}
\hline
Benchmark Points & $\sigma_{4b}$ & $\sigma_{4bl}$ &$\sigma_{2b2l}$\\
\hline
\hline
BP1 & 0.115 & 0.013 & 0.029 \\
\hline
BP2 & 0.080 & 0.0129 & 0.0423 \\
\hline
BP3 & 0.0765 & 0.0099 & 0.0234 \\
\hline
MSG1 & 0.0017 & 0.00062 & 0.0311 \\
\hline
ttbar & 0.0000 & 0.0000 & 0.0420 \\
\hline
\hline
\end {tabular}
\end{center}
{\caption {\em Event-rates (pb) in multichannels at the chosen
benchmark points for $E_{CM}$= 7 TeV. {\tt CTEQ5L} pdfset was
used. Factorization and Renormalization scale has been set to
$\mu_F=\mu_R=\sqrt{\hat s}$, subprocess centre of mass energy.}}
\end{table}

We also show
distributions in different channels for (i) Missing Energy, (ii) Sum
over $p_T$ of $b$s, leptons with Missing energy. The last one is
really close to the definition of so-called Effective Mass. We show these distributions at 14 TeV. Missing energy distribution mostly indicate the LSP mass of the underlying benchmark points. This distribution peaks around 2$m_{\tilde{{\chi_1}^0}}$. Hence, the flatter distribution is for MSG1 among the benchmarks. On the other hand, the effective mass distribution picks out the mass of the dominant production process. The peak of the distribution is around $2m_{\tilde{g}}$. Here as well, MSG1 shows the flattest distribution among others. While these distributions show the robustness of our analysis, it is also true that this can't be used really to distinguish these models, which have masses quite near to each other. However, that is not our intention really in this analysis.

\begin{figure}[htbp]
\begin{center}
\centerline{\psfig{file=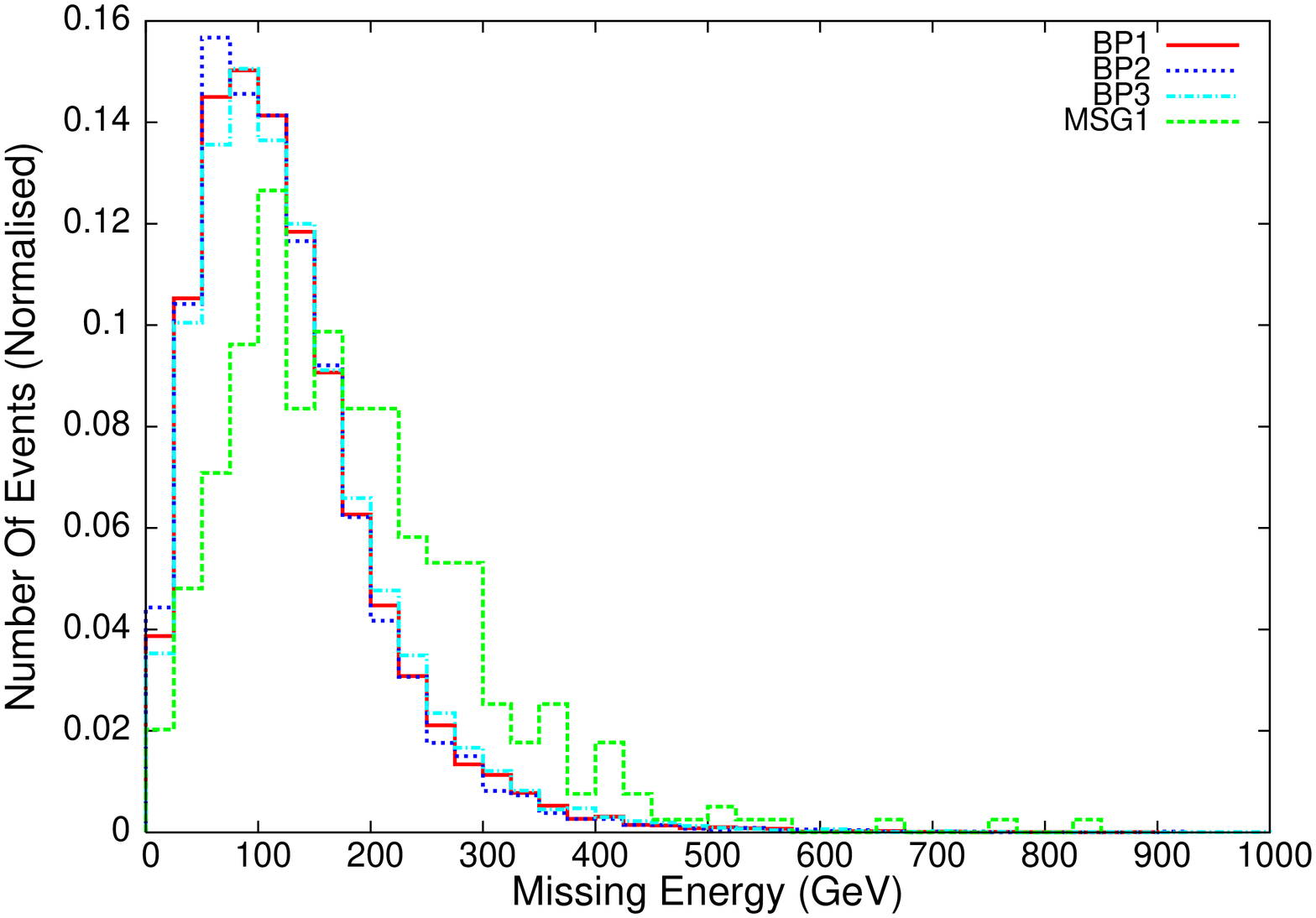,width=6.5 cm,height=7.5cm,angle=-90}
\hskip 20pt \psfig{file=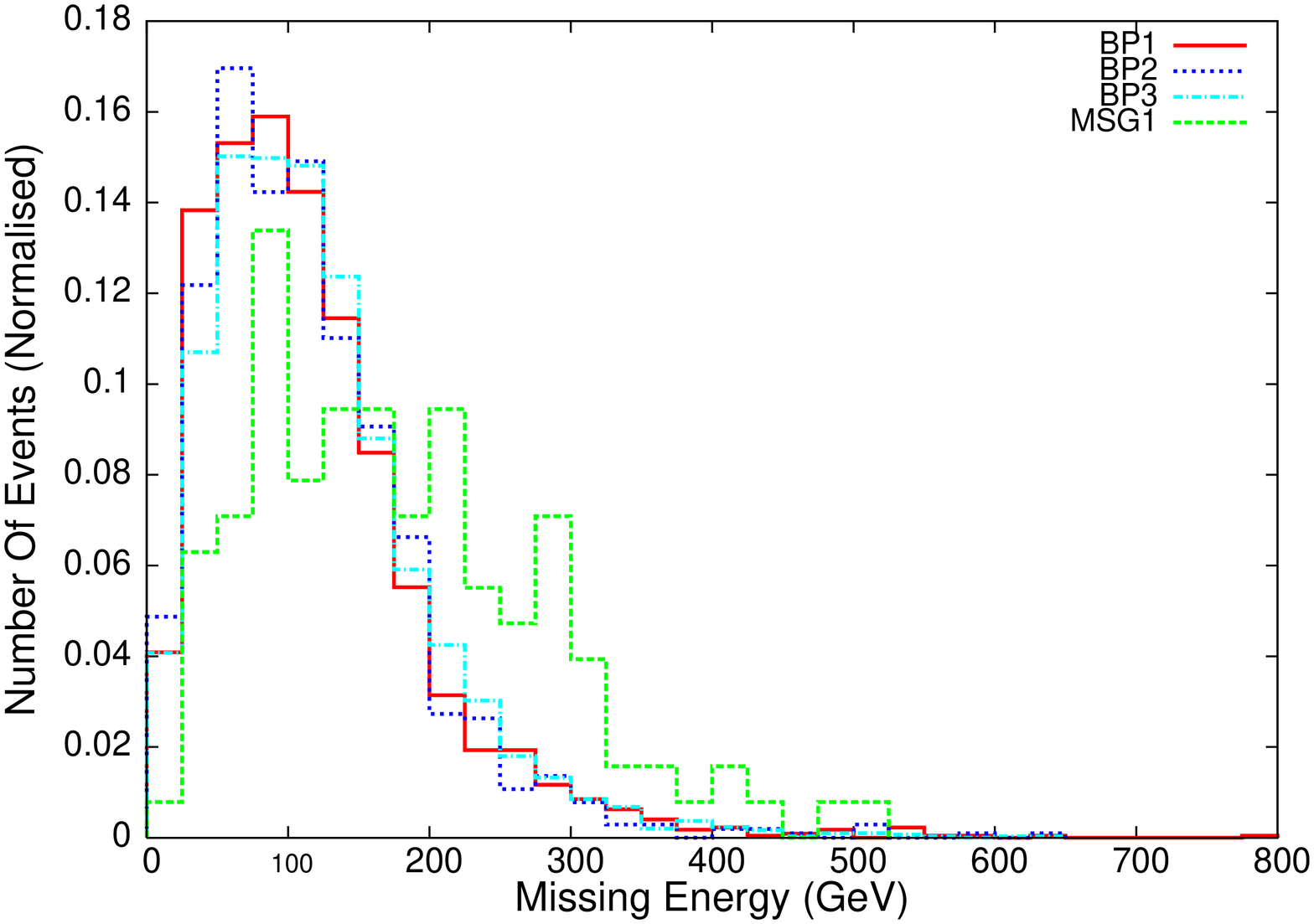,width=6.5 cm,height=7.5cm,angle=-90}}
\vskip 10pt
\centerline{\psfig{file=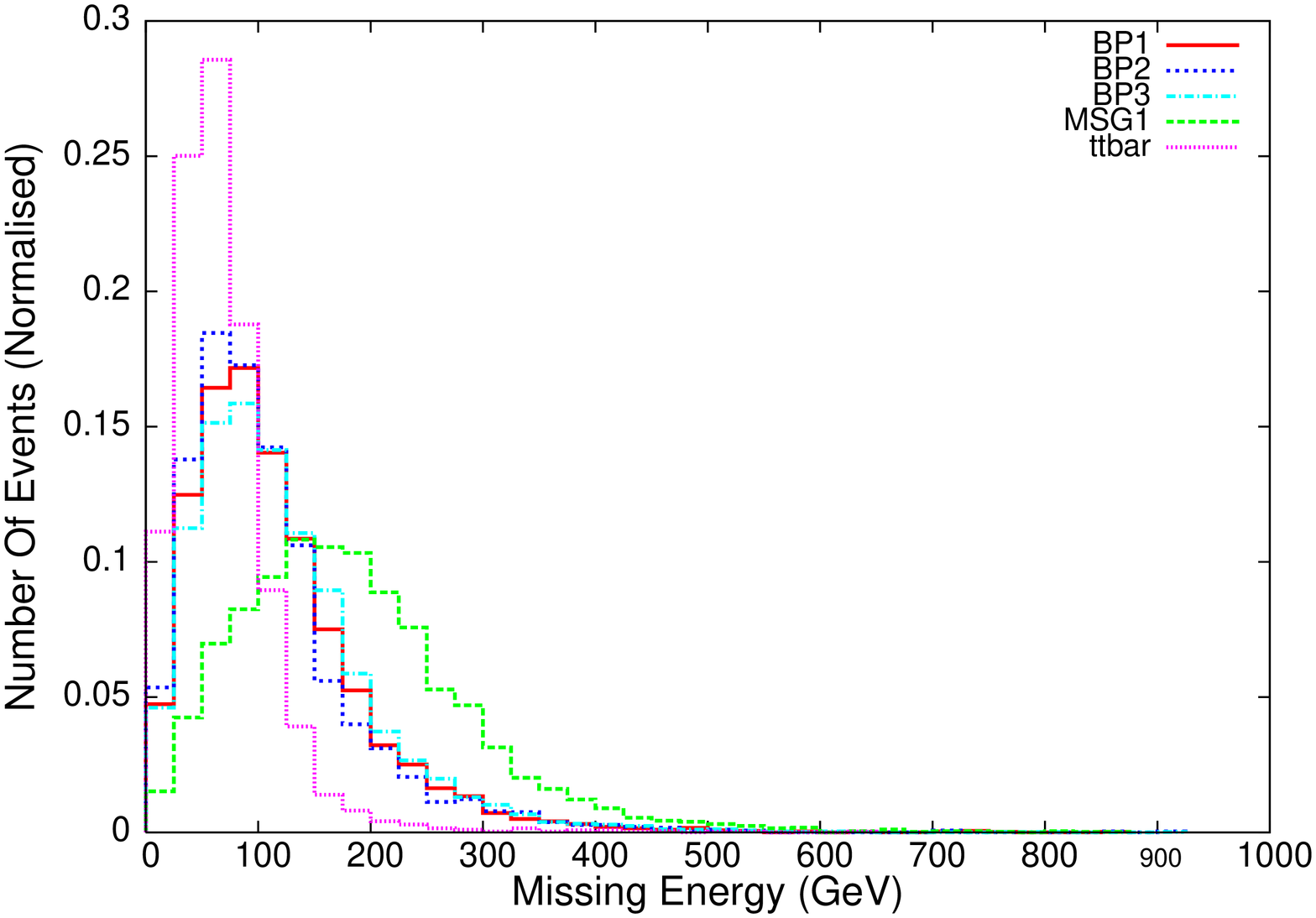,width=6.5 cm,height=7.5cm,angle=-90}}
\caption{{\em Missing energy distribution in $4b$, $4bl$ and $2b2l$ events 
from left to right and top to bottom before putting cut. 
{\tt CTEQ5L} pdfset was used. Factorization and 
Renormalization scale has been set to $\mu_F=\mu_R=\sqrt{\hat s}$, 
sub-process centre of mass energy.}} 
\end{center}
\end{figure}

\begin{figure}[htbp]
\begin{center}
\centerline{\psfig{file=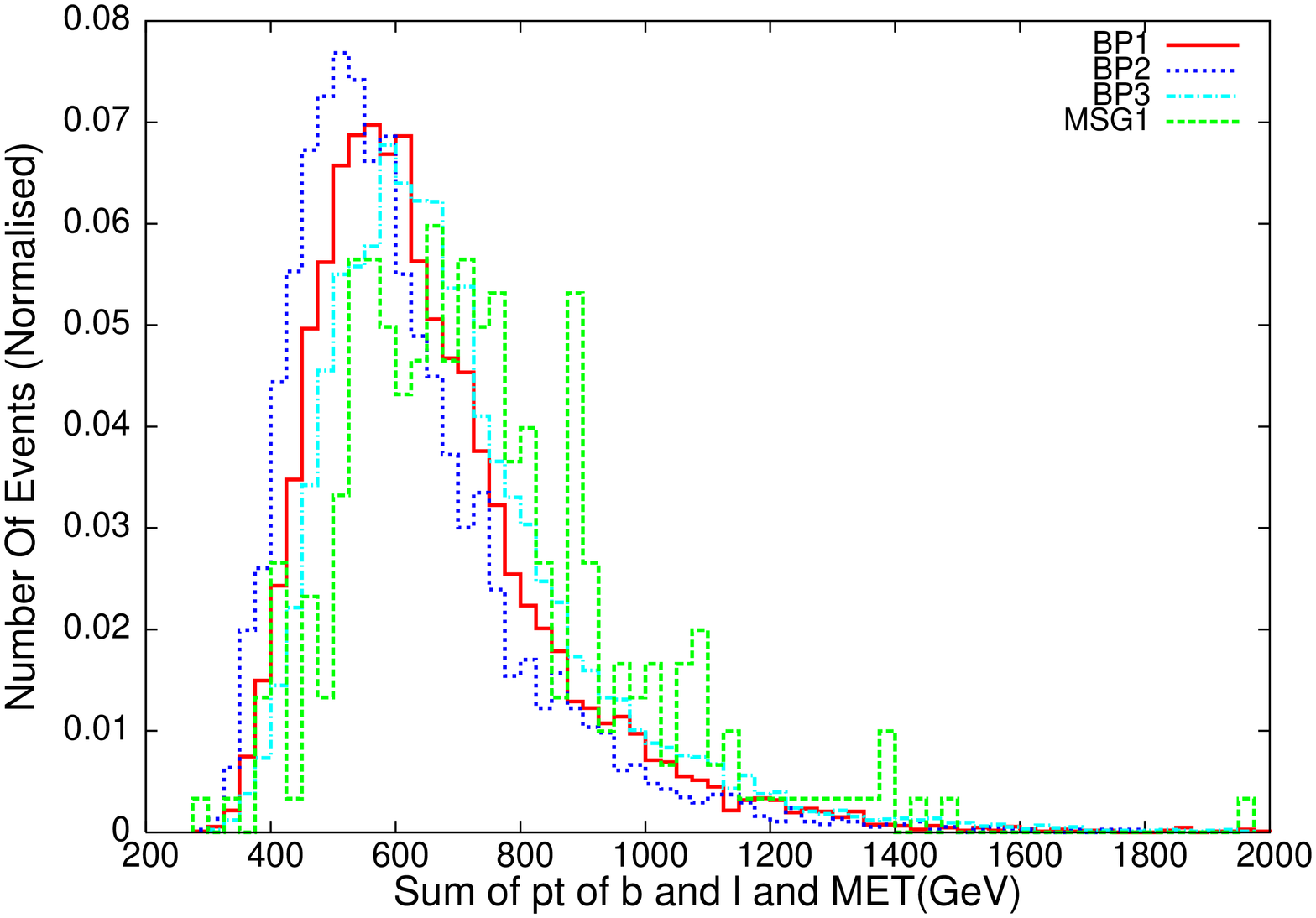,width=6.5
cm,height=7.5cm,angle=-90} 
\hskip 20pt
\psfig{file=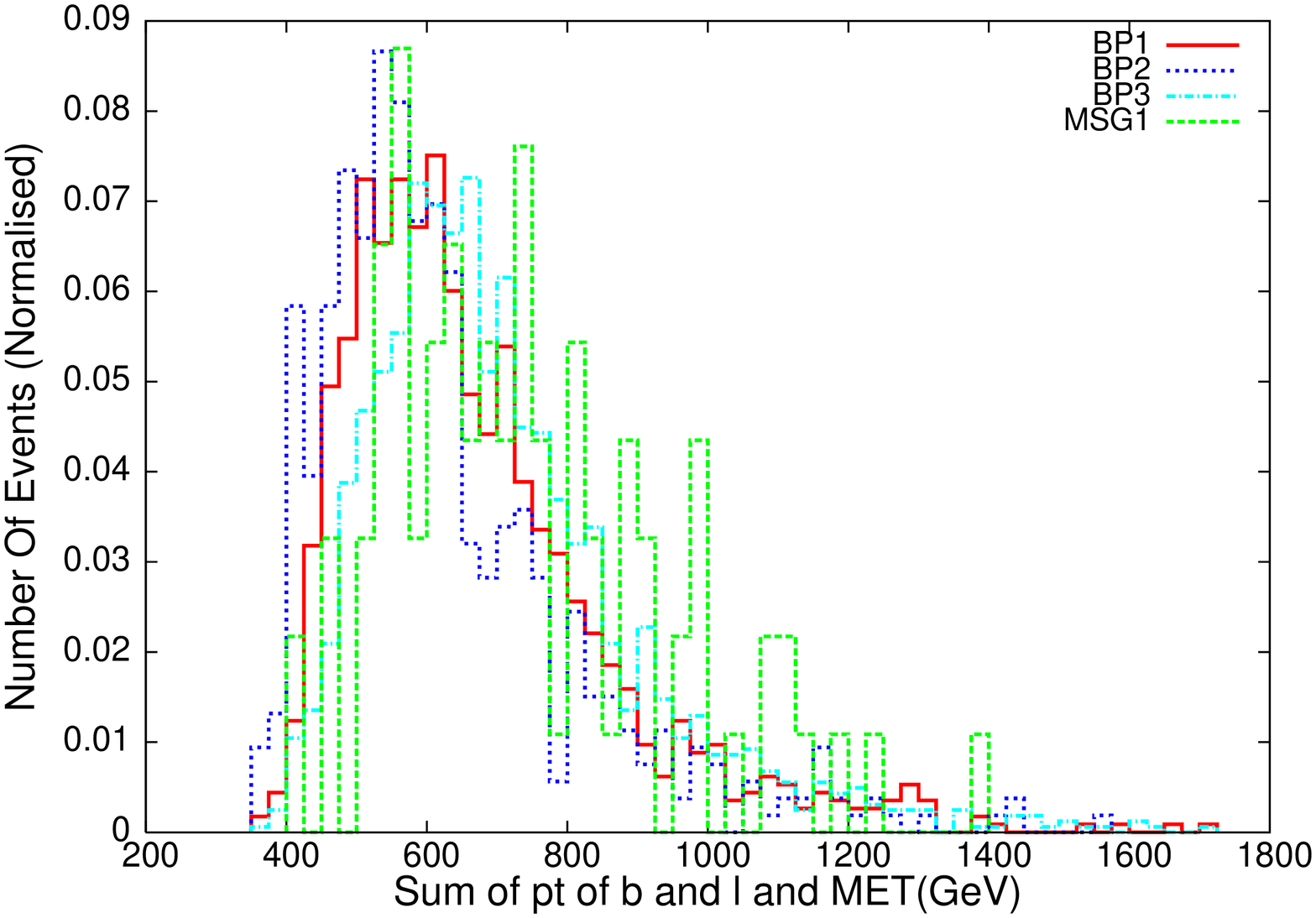,width=6.5 cm,height=7.5cm,angle=-90}}
\vskip 10pt 
\centerline{\psfig{file=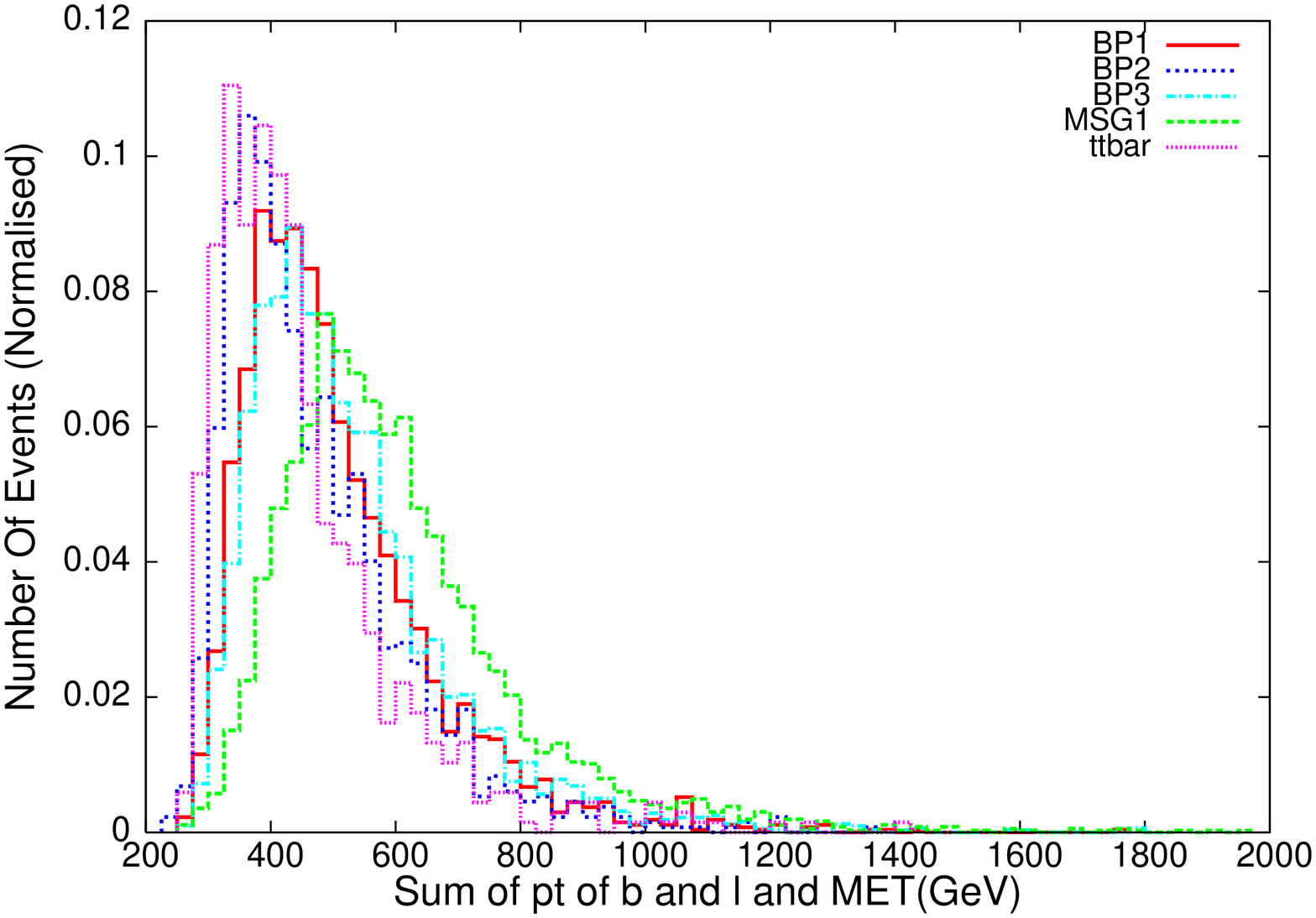,width=6.5
cm,height=7.5cm,angle=-90}} 
\caption{{\em `$\Sigma p_T$ of $b$s and
leptons and missing energy' distribution in $4b$, $4bl$ and $2b2l$
events from left to right and top to bottom before putting cut. {\tt
CTEQ5L} pdfset was used. Factorization and Renormalization scale has
been set to $\mu_F=\mu_R=\sqrt{\hat s}$, sub-process centre of mass
energy.}}
\end{center}
\end{figure}

\section {Summary and Conclusions}

{In this work, we have proposed a scenario in which discovery of supersymmetry may be possible at the early runs of the LHC. Our scenario includes a light gluino and light 3rd family squarks, whereas the squarks for the first two families  are very heavy. We work within the gravity mediated SUSY breaking scenario, but go beyond mSUGRA using non-universal boundary conditions for the gaugino masses, as well as for the sfermion masses.The gluinos are light enough to be pair produced copiously at the LHC. Their decays to sbottom-botom and stop-top
gives rise to bottom quark rich final states with or without charged leptons together with large missing energy associated with the lightest (LSP) neutralino. The most promising final states are
$4b+{E_{T}}\!\!\!\!/$, $4b+\ell+{E_{T}}\!\!\!\!/$ and
$2b+2\ell+{E_{T}}\!\!\!\!/$. We have chosen three benchmark points and also one mSUGRA point
with light gluino mass for comparison. With the cuts appropriate for the LHC, we find that the signal cross section for the $4b+{E_{T}}\!\!\!\!/$ channel is well above $1,000$ fb, whereas for the other two channels, it is well above $100$ fb. For comparison, the signal for the mSUGRA point (MSG1) is much smaller.  The dominant background with the used cuts are from $t\bar{t}$ production which is significant for the $2b+2\ell+{E_{T}}\!\!\!\!/$ channel, but negligible for the $4b+{E_{T}}\!\!\!\!/$, $4b+\ell+{E_{T}}\!\!\!\!/$ channels. Thus as the LHC accumulates data this year to few hundred $pb^{-1}$  (or $1 fb^{-1}$), the prospect of discovering SUSY in our proposed scenario is excellent., and we urge the ATLAS and CMS Collaborations to search for SUSY in these final states.}


{ NOTE ADDED: During the completion of writing of this manuscript an article  \cite{Kane}
in similar direction appeared in the arXiV. These authors also emphasize similar final states as in our work. They do not specify any SUSY breaking model, but rather use three different branching ratio for the gluino decays to choose their benchmark points. In our work, we use gravity mediated SUSY breaking framework, and generate the SUSY spectra from the high scale  with non-universal boundary conditions, and satisfy the experimental constraints on the SUSY particle masses , dark matter and other low energy processes. Our results for the cross sections for the final states are in good agreement with their results.}

{\bf Acknowledgment:} SB would like to thank SN for his invitation to visit OSU and for the  hospitality provided during the stay, when this work was formulated and carried out to a large extent. SB also acknowledges the support from Dr. Khanov at OSU for providing computer and cluster access which was used to a large extent for the computational part of this work.  





\end{document}